\documentclass[fleqn,twoside]{article}
\usepackage[headings]{espcrc2}

\readRCS
$Id: espcrc2.tex,v 1.2 2004/02/24 11:22:11 spepping Exp $
\usepackage{graphicx}
\usepackage[figuresright]{rotating}

\newcommand{\AmS}{{\protect\the\textfont2
  A\kern-.1667em\lower.5ex\hbox{M}\kern-.125emS}}

\usepackage{axodraw}
\usepackage{feynarts}
\usepackage{color}

\def\beq{\begin{equation}}
\def\eeq{\end{equation}}
\def\beqar{\begin{eqnarray}}
\def\eeqar{\end{eqnarray}}
\def\barr#1{\begin{array}{#1}}
\def\earr{\end{array}}
\def\bfi{\begin{figure}}
\def\efi{\end{figure}}
\def\btab{\begin{table}}
\def\etab{\end{table}}
\def\bce{\begin{center}}
\def\ece{\end{center}}
\def\nl{\nonumber\\*}
\def\nn{\nonumber}

\def\text{\textstyle}

\newcommand{\rd}{\mathrm{d}}
\newcommand{\ri}{\mathrm{i}}

\def\de{\delta}

\newcommand{\ina}{i_1}
\newcommand{\inb}{i_2}
\newcommand{\inc}{i_3}

\newcommand{\Zadj}{\tilde Z}
\newcommand{\Zadjadj}{\,\smash{\tilde{\!\tilde Z}}\vphantom{\tilde Z}}
\newcommand{\Ymod}{X}
\newcommand{\Ymodadj}{\tilde{\Ymod}}

\newcommand{\Gramdet}{|Z|}

\renewcommand{\Green}{\textcolor{green}}

\definecolor{mygreen}{rgb}{0,1,0}
\definecolor{mygreen}{rgb}{0,.65,0}
\renewcommand{\Green}{\textcolor{mygreen}}
\definecolor{mycyan}{cmyk}{1,0,0,0}
\definecolor{mycyan}{cmyk}{.8,.25,0,0}

\definecolor{mymagenta}{cmyk}{0,1,0,0}
\definecolor{mymagenta}{cmyk}{.25,1,0,0}

\def\mathswitchr#1{\relax\ifmmode{\mathrm{#1}}\else$\mathrm{#1}$\fi}

\newcommand{\PW}{\mathswitchr W}
\newcommand{\PZ}{\mathswitchr Z}

\newcommand{\Pe}{\mathswitchr e}

\newcommand{\Pd}{\mathswitchr d}
\newcommand{\Pu}{\mathswitchr u}

\newcommand{\Pep}{\mathswitchr {e^+}}
\newcommand{\Pem}{\mathswitchr {e^-}}

\def\mathswitch#1{\relax\ifmmode#1\else$#1$\fi}

\newcommand{\MZ}{\mathswitch {M_\PZ}}

\newcommand{\Mu}{\mathswitch {m_\Pu}}

\newcommand{\GeV}{\unskip\,\mathrm{GeV}}

\def\draftdate{\relax}
\def\mda{\relax}
\def\mua{\relax}
\def\mla{\relax}
\def\draft{
\def\thtystars{******************************}
\def\sixtystars{\thtystars\thtystars}
\typeout{}
\typeout{\sixtystars**}
\typeout{* Draft mode!
         For final version remove \protect\draft\space in source file *}
\typeout{\sixtystars**}
\typeout{}
\def\draftdate{\today}
\def\mua{\marginpar[\boldmath\hfill$\uparrow$]%
                   {\boldmath$\uparrow$\hfil}%
                    \typeout{marginpar: $\uparrow$}\ignorespaces}
\def\mda{\marginpar[\boldmath\hfill$\downarrow$]%
                   {\boldmath$\downarrow$\hfil}%
                    \typeout{marginpar: $\downarrow$}\ignorespaces}
\def\mla{\marginpar[\boldmath\hfill$\rightarrow$]%
                   {\boldmath$\leftarrow $\hfil}%
                    \typeout{marginpar: $\rightarrow$}\ignorespaces}
\def\Mua{\marginpar[\boldmath\hfil$\Uparrow$]%
                   {\boldmath$\Uparrow$\hfil}%
                    \typeout{marginpar: $\uparrow$}\ignorespaces}
\def\Mda{\marginpar[\boldmath\hfil$\Downarrow$]%
                   {\boldmath$\Downarrow$\hfil}%
                    \typeout{marginpar: $\downarrow$}\ignorespaces}
\def\Mla{\marginpar[\boldmath\hfil$\Rightarrow$]%
                   {\boldmath$\Leftarrow $\hfil}%
                    \typeout{marginpar: $\leftarrow$}\ignorespaces}
\overfullrule 5pt
\marginparwidth 20mm
}

\title{Techniques for one-loop tensor integrals in many-particle processes}

\author{A.~Denner\address{Paul Scherrer Institut, W\"urenlingen und Villigen,
        CH-5232 Villigen PSI, Switzerland}
        and
        S.~Dittmaier\address{Max-Planck-Institut f\"ur Physik
        (Werner-Heisenberg-Institut),
        D-80805 M\"unchen, Germany}}
       
\runtitle{Techniques for one-loop tensor integrals in many-particle processes}
\runauthor{A.~Denner and S.~Dittmaier}

\begin{document}

\begin{abstract}
We briefly sketch the methods for a numerically stable evaluation of
tensor one-loop integrals that have been used in the calculation of the 
complete electroweak one-loop corrections to $\Pep\Pem\to4\,$fermions.
In particular, the improvement of the new methods over the conventional
Passarino--Veltman reduction is illustrated for some 4-point integrals
in the delicate limits of small Gram (and other kinematical)
determinants.
\vspace{1pc}
\end{abstract}

\maketitle

\vspace*{-4em}
\section{INTRODUCTION}

At the LHC and ILC, many interesting processes involve more than four
external particles. Such many-particle
reactions often proceed via resonances that subsequently
decay, or they represent a background to resonance processes.
A thorough description of such processes requires the evaluation
of strong and electroweak radiative corrections at least in
next-to-leading order. The most complicated part in such calculations
concerns the numerically stable evaluation of the one-loop tensor
integrals of the virtual corrections. 

For processes with up to four external particles the classical 
Passarino--Veltman (PV) reduction \cite{Passarino:1978jh}, which
recursively reduces tensor to scalar integrals,
is sufficient in practically all cases. 
This scheme, however, involves
Gram determinants in the denominator,
which spoil the numerical stability if they become small. 
With up to four external particles this happens only near the
edge of phase space (forward scattering, thresholds). 
With more than four external particles, Gram determinants
also vanish within phase space, and methods 
are needed where Gram determinants can 
be small but still non-zero.
Several solutions to this problem have been proposed in recent years,
but not many of them have proven their performance in complicated
applications yet. 
For references and descriptions of some methods 
alternative to ours, we refer to Refs.~\cite{Denner:2005nn,LH2005}.

In this article 
we briefly describe the methods used in the calculation of the 
complete electroweak ${\cal O}(\alpha)$ corrections to $\Pep\Pem\to4\,$fermions
\cite{Denner:2005es}, which constitutes the first established one-loop result
for a $2\to4$ particle reaction.
All relevant formulas can be found in Refs.~\cite{Denner:2005nn,Denner:2002ii},
here we only sketch their structure.
Moreover, we illustrate the improvement of the new methods over the conventional
PV reduction for some 4-point integrals
in delicate kinematical limits.

\section{THE GENERAL CONCEPT}

One-loop tensor integrals can be naturally grouped into three categories,
which we have treated in completely different ways:

(i)
For {\it 1- and 2-point integrals} of arbitrary tensor rank, 
numerically stable analytical expressions are presented in 
Ref.~\cite{Denner:2005nn} (see also Ref.~\cite{Passarino:1978jh}).

(ii)
For {\it 3- and 4-point tensor integrals},
PV reduction \cite{Passarino:1978jh}
is applied for ``regular'' phase-space points where Gram determinants
are not too small. For the remaining problematic cases special reduction
techniques have been developed~\cite{Denner:2005nn}.

One of the techniques replaces the
standard scalar integral by a specific tensor coefficient that can be
safely evaluated numerically and reduces the remaining tensor
coefficients as well as the standard scalar integral to the new 
basis integrals. In this scheme no dangerous inverse Gram determinants
occur, but inverse modified Cayley determinants instead. 
The procedure is related to the fully numerical method described in
Ref.~\cite{Ferroglia:2002mz}.

In a second class of techniques, 
the tensor coefficients are iteratively deduced up to terms that are
systematically suppressed by small Gram or other
kinematical determinants in specific kinematical configurations. The
numerical accuracy can be systematically improved upon
including higher tensor ranks.
A similar idea, where tensor coefficients are iteratively determined
from higher-rank tensors has been described in Ref.~\cite{Giele:2004ub}
for the massless case.

These methods are sketched in more detail and illustrated in
4-point examples below.

(iii)
For {\it 5- and 6-point integrals}, direct
reductions to 5- and 4-point integrals, respectively, are possible
owing to the four-dimensionality of space-time. 
For scalar integrals such a reduction was already derived in the 
1960s \cite{Me65}.
In Refs.~\cite{Denner:2005nn,Denner:2002ii} we follow basically the same
strategy to reduce tensor integrals, which has the advantage that 
no inverse Gram determinants appear in the reduction.
Instead 
modified Cayley determinants occur in the denominator,
but we did not find numerical problems with these factors.
A reduction similar to ours
has been proposed in Ref.~\cite{Binoth:2005ff}.
\looseness-1

\vspace{1em}
We would like to stress two important features of our approach.

(i)
The methods are valid for massive and massless cases. 
The formulas given in Refs.~\cite{Denner:2005nn,Denner:2002ii} are valid
without modifications if IR divergences are regularized with mass
parameters or dimensionally.%
\footnote{For the method of Ref.~\cite{Denner:2002ii}, this has been
shown in Ref.~\cite{Dittmaier:2003bc}.}
Finite masses can be either real or complex.

(ii)
The in/out structure of the methods is the same as for
conventional PV reduction, i.e.\ 
no specific algebraic manipulations are needed in applications.
Therefore, the whole method can be (and in fact is) organized as a numerical
library for scalar integrals and tensor coefficients.

\vspace{1em}
We conclude this overview with some comments resulting from our
experience collected in the treatment of a full $2\to4$ scattering
reaction. 

(i) For a specific point in a multi-particle (multi-parameter) phase
space it is highly non-trivial to figure out which of the various
methods is the most precise.  It seems hopeless to split the phase
space into regions that are dedicated to a given method.  Therefore,
we estimate the accuracy for the different methods at each
phase-space point and take the variant promising the highest
precision. In one approach, we estimate
the number of valid digits based on the expected accuracy of the
expansions and possible numerical cancellations before the evaluation
of the coefficients. 
In a second approach, the accuracy of the PV method is estimated 
by checking symmetries and sum rules for the coefficients.
If the estimated precision is not
satisfactory, the seminumerical method is taken; if this is still
not satisfactory, an expansion is used.

(ii)       
In a complicated phase space it may happen that none of the
various methods is perfect or good in some exceptional situations.
Usually the corresponding events 
do not significantly contribute to cross sections. 
This issue can only be fathomed in actual applications. 
To be on the safe side, we employ the two independent ``rescue systems''
with different advantages and limitations.

(iii)
In view of this, figures as shown below are nice
illustrations, but should always be taken with a grain of salt.
No matter how many of such figures are shown, they will never be
exhaustive, so that no quantitative conclusions on the overall
precision of methods can be drawn.

In summary, the performance
(speed, stability, etc.) of any method can only be estimated and
proven in non-trivial and realistic applications.

\section{4-POINT INTEGRALS FOR SMALL GRAM DETERMINANTS}

In the following we sketch the methods of Ref.~\cite{Denner:2005nn}
for the 4-point tensor integrals of rank~$P$,
\beqar
\lefteqn{
D^{\mu_{1}\ldots\mu_{P}}
=\frac{(2\pi\mu)^{4-D}}{\ri\pi^{2}}\int \rd^{D}q\,
\frac{q^{\mu_{1}}\cdots q^{\mu_{P}}}
{N_0N_1N_2N_3},
}
\nn\\ && {}
N_k= (q+p_k)^2-m_k^2+\ri\epsilon, \quad p_0=0,
\eeqar
which are decomposed into covariants as follows,
\beqar
\lefteqn{
D^{\mu}=\sum_{\ina=1}^{3} p_{\ina}^{\mu}D_{\ina},
}
\nn\\ 
\lefteqn{
D^{\mu\nu}=\sum_{\ina,\inb=1}^{3} p_{\ina}^{\mu}p_{\inb}^{\nu}D_{\ina\inb}
+g^{\mu\nu}D_{00},
}
\nn\\ 
\lefteqn{
D^{\mu\nu\rho}=\sum_{\ina,\inb,\inc=1}^{3} p_{\ina}^{\mu}p_{\inb}^{\nu}p_{\inc}^{\rho}D_{\ina\inb\inc}
}
\nn\\ && {}
+\sum_{\ina=1}^{3}
(g^{\mu\nu}p_{\ina}^{\rho}+g^{\nu\rho}p_{\ina}^{\mu}+g^{\rho\mu}p_{\ina}^{\nu})
D_{00\ina},
\eeqar
and so on for higher rank.
The conventional PV reduction \cite{Passarino:1978jh}
recursively expresses the tensor coefficients
$D_{i_1\dots i_P}$ in terms of the scalar 4-point integral $D_0$ and
3-point integrals of lower rank. 
To this end, the defining tensor
integrals are contracted with an external momentum $p^{\mu_1}_k$
or with the metric tensor $g^{\mu_1 \mu_2}$, leading to the factors
$p_k q$ or $q^2$ in the numerator, respectively. Rewriting these factors
according to
\beqar
2p_k q &=& N_k-N_0-f_k, \quad
f_k=p_k^2-m_k^2+m_0^2,\nl
q^2 &=& N_0 + m_0^2,
\eeqar
the $N_i$ terms can be used to cancel propagator denominators.
This procedure yields the basic PV relations that
connect contracted 4-point tensor integrals of rank $P$ with
lower-rank 4- and 3-point integrals.
The PV solution expresses the rank $P$ 4-point
coefficients in terms of lower-rank 4- and 3-point coefficients.
In each step $P\to(P-1)$
the inverse of the Gram matrix
\beq
Z = \left(\barr{ccc}
2 p_1 p_1 & 2 p_1 p_2 & 2 p_1 p_3 \\
2 p_2 p_1 & 2 p_2 p_2 & 2 p_2 p_3 \\
2 p_3 p_1 & 2 p_3 p_2 & 2 p_3 p_3
\end{array} \right)
\eeq
occurs, which causes the above-mentioned numerical problems if the 
determinant $|Z|$ becomes small.
In our alternative methods also the matrix
\beqar
\lefteqn{
X = \left(
\barr{c|ccc}
\! 2m_0^2 & \! f_1 & \!\! f_2 & \!\! f_3 \! \\
\hline
\barr{c} f_1 \\ f_2 \\ f_3 \earr 
& & Z & \\
\earr\right), 
}
\nn\\[-1.5em]
\eeqar
and its inverse appear. The vanishing
of the modified Cayley determinant $|X|$ corresponds
to necessary conditions
for true (Landau) singularities in a Feynman diagram.

We adopt the conventions that indices $\hat i$ with a hat
should be omitted and that $\Zadj_{ij}$ and $\Ymodadj_{ij}$ denote
the minors (i.e.\ determinants of submatrices where row $i$ and column
$j$ are discarded) of the matrices $Z$ and $X$, respectively. 
Similarly $\Zadjadj_{ij,kl}$ denotes generalized minors where    
two rows and two columns are discarded in $Z$. For the precise
definitions, full formulas, and detailed descriptions of
algorithms we refer to Ref.~\cite{Denner:2005nn}.
Here we only highlight the general structure.

\subsection{\boldmath{Seminumerical method for small $|Z|$}}

The PV relations can be arranged as follows,
\beqar
X \left(\barr{c}
\hspace*{-.5em} D_{i_2\ldots i_P}  \hspace*{-.5em}\\
\hspace*{-.5em} D_{1i_2\ldots i_P} \hspace*{-.5em}\\
\hspace*{-.5em} D_{2i_2\ldots i_P} \hspace*{-.5em}\\
\hspace*{-.5em} D_{3i_2\ldots i_P} \hspace*{-.5em}\\
\earr\right)
&\hspace*{-.5em}=\hspace*{-.5em}& -2\left(\barr{cccc}
\hspace*{-.5em}
\mbox{\small const.}\times {D_{00i_2\ldots i_P}} \hspace*{-.5em}\\
\hspace*{-.5em}\sum_{r=2}^P \de_{1i_r} {D_{00i_2\ldots \hat i_r\ldots i_P}} \hspace*{-.5em}\\
\hspace*{-.5em}\sum_{r=2}^P \de_{2i_r} {D_{00i_2\ldots \hat i_r\ldots i_P}} \hspace*{-.5em}\\
\hspace*{-.5em}\sum_{r=2}^P \de_{3i_r} {D_{00i_2\ldots \hat i_r\ldots i_P}} \hspace*{-.5em}\\
\earr\right)
\nn\\
&& {}
+ \mbox{\small 3-point integrals,} 
\eeqar
which expresses tensor coefficients of rank $P$ and $(P-1)$ in terms
of coefficients of rank $P$ and $(P+1)$ that involve one metric tensor
more in the corresponding covariants.
Thus, these relations can be used to recursively express all 4-point
coefficients up to rank $P$ (including the scalar integral)
in terms of the coefficient 
$D_{0\dots0}$ of rank $2P$ and 3-point integrals.
For rank~4 or higher the coefficient $D_{0\dots0}$ has the property
that its Feynman-parameter integral involves a logarithmic integrand
without any denominator and is, thus, well suited for numerical
integration.

\begin{figure*}
\hspace*{6em}
{\scriptsize\unitlength0.7pt 
\begin{picture}(150,95)(0,-50)
\SetScale{.7}
\put(-40,53){\small Full diagram:}
\ArrowLine(40,20)(10,20)
\ArrowLine(10,-20)(40,-20)
\ArrowLine(130,15)(100,15)
\ArrowLine(80,40)(130,40)
\ArrowLine(100,-15)(130,-15)
\ArrowLine(130,-40)(80,-40)
\SetColor{PineGreen}
\ArrowLine(100,15)(80,40)
\ArrowLine(40,-20)(40,20)
\ArrowLine(80,-40)(100,-15)
\Photon(40,20)(80,40){2}{4.5}
\SetColor{Black}
\Photon(100,-15)(100,15){2}{3.5}
\Photon(40,-20)(80,-40){-2}{4.5}
\Vertex(100,-15){2}\Vertex(100,15){2}
\Vertex(40,20){2}\Vertex(80,40){2}
\Vertex(40,-20){2}\Vertex(80,-40){2}
\Text(6,20)[r]{$\Pe^+$}
\Text(6,-20)[r]{$\Pe^-$}
\Text(35,0)[r]{$\Pe$}
\Text(135,40)[l]{$\mu^-$}
\Text(135,15)[l]{$\bar\nu_\mu$}
\Text(135,-15)[l]{$\Pu$}
\Text(135,-40)[l]{$\bar\Pd$}
\Text(60,40)[b]{$\PZ$}
\Text(60,-40)[t]{$\PZ$}
\Text(105,0)[l]{$\PW$}
\Text(98,27)[l]{$\mu$}
\Text(98,-27)[l]{$\Pd$}
\end{picture}}
\hspace*{8em}
{\scriptsize \unitlength .6pt
\begin{picture}(210,120)(10,-15)
\put(-35,105){\small Subdiagram:}
\SetScale{.6}
\Line(20, 85)( 60, 85)
\Line(20,  5)( 60,  5)
\Line(140, 85)(180, 85)
\Line(180,  5)(140,  5)
\SetColor{PineGreen}
\Line(60, 85)(140, 85)
\Line(60,  5)(140,  5)
\Line(60, 85)( 60,  5)
\Line(140,  5)(140, 85)
\SetColor{Black}
\Vertex( 60, 85){3}
\Vertex( 60,  5){3}
\Vertex(140,  5){3}
\Vertex(140, 85){3}
\put(64,40){$0$}
\put(95,11){$0$}
\put(92,67){$\MZ$}
\put(126,40){$0$}
\put( 6,80){$0$}
\put( 0,-5){$t_{\Pe\bar\Pd}$}
\put(185,80){$0$}
\put(185,-5){$s_{\bar\nu\Pu}$}
\CArc(100,50)(50,60,120)
\put(90,110){$t_{\bar\Pe\mu}$}
\CArc(110,45)(50,-30,30)
\put(165,40){$s_{\mu\bar\nu\Pu}$}
\end{picture} }
\\[.5em]
\includegraphics[scale=0.7]{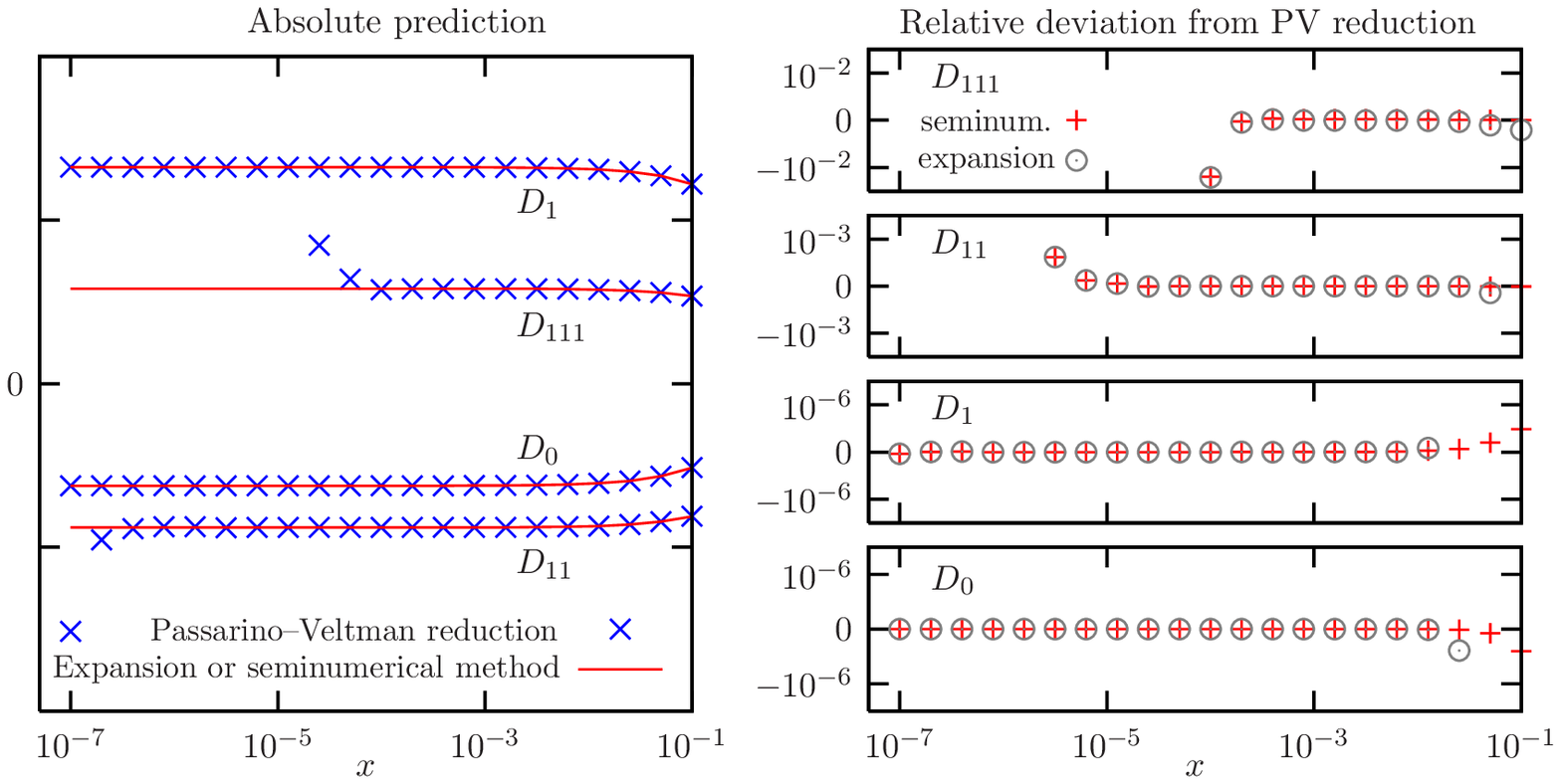}
\hspace*{0em}
\raisebox{3em}{\unitlength 1cm
\begin{picture}(3.7,4.2)
\put(0,4.0){\small Kinematics:}
\put(0.1,3.4){\scriptsize \parbox{.7cm}{$s_{\mu\bar\nu\Pu}$}$=+2{\times}10^4\GeV^2$}
\put(0.1,3.0){\scriptsize \parbox{.7cm}{$s_{\bar\nu\Pu}$}$=+1{\times}10^4\GeV^2$}
\put(0.1,2.6){\scriptsize \parbox{.7cm}{$t_{\bar\Pe\mu}$}$=-4{\times}10^4\GeV^2$}
\put(0.1,2.0){\scriptsize \parbox{.7cm}{$t_{\mathrm{crit}}$}$= \displaystyle
\frac{s_{\mu\bar\nu\Pu}(s_{\mu\bar\nu\Pu}-s_{\bar\nu\Pu}+t_{\bar\Pe\mu})}
{s_{\mu\bar\nu\Pu}-s_{\bar\nu\Pu}}$}
\put(0.1,1.4){\scriptsize \parbox{.7cm}{\mbox{}}$=-6{\times}10^4\GeV^2$}
\put(0.1,0.6){$|Z|\to0 \;\Leftrightarrow$}
\put(0.1,0.1){{$x\equiv {t_{\Pe\bar\Pd}}/{t_{\mathrm{crit}}}-1\to0$}}
\end{picture} }
\vspace{-3.5em}
\caption{A typical example for 4-point integrals with small
$|Z|$ ($x\to0$). 
The full diagram and the relevant subdiagram
are given above; absolute predictions (in arbitrary units) for some tensor 
coefficients, relative deviations from PV reduction, and
the kinematic specifications are shown below.}
\label{fig:DsmallGram}
\vspace*{1em}
\hspace*{6em}
\raisebox{-1em}{\scriptsize
\unitlength=1.1bp%
\begin{feynartspicture}(100,80)(1,1)
\FADiagram{}
\FAProp(6.,15.)(6.,5.)(0.,){/Straight}{-1}
\FAProp(0.,15.)(6.,15.)(0.,){/Straight}{-1}
\FALabel(3.,16.27)[b]{$\Pep$}
\FAProp(0.,5.)(6.,5.)(0.,){/Straight}{1}
\FAProp(20.,15.)(14.,15.)(0.,){/Straight}{-1}
\FALabel(3.,3.93)[t]{$\Pem$}
\FALabel(17.,16.07)[b]{$\Pu$}
\FAProp(20.,5.)(14.,5.)(0.,){/Straight}{1}
\FALabel(17.,3.73)[t]{$\bar\Pd$}
\FAProp(20.,10.)(14.,10.)(0.,){/Sine}{-1}
\FALabel(17.,11.07)[b]{$\PW$}
\FALabel(4.93,10.)[r]{$\Pe$}
\FALabel(10.,16.27)[b]{$\gamma$}
\Green{
\FAProp(6.,15.)(14.,15.)(0.,){/Sine}{0}
\FAProp(6.,5.)(14.,5.)(0.,){/Sine}{0}
\FAProp(14.,15.)(14.,10.)(0.,){/Straight}{-1}
\FAProp(14.,5.)(14.,10.)(0.,){/Straight}{1}
}
\FALabel(10.,3.93)[t]{$\PZ$}
\FALabel(12.73,12.5)[r]{$\Pu$}
\FALabel(12.73,7.5)[r]{$\Pd$}
\FAVert(6.,15.){0}
\FAVert(6.,5.){0}
\FAVert(14.,15.){0}
\FAVert(14.,5.){0}
\FAVert(14.,10.){0}
\FAVert(20.,10.){0}
\FAProp(20.,10.)(24.,12.)(0.,){/Straight}{1}
\FAProp(24.,8.)(20.,10.)(0.,){/Straight}{1}
\FALabel(26,13.5)[t]{$\mu^-$}
\FALabel(26,8)[t]{$\bar\nu_\mu$}
\end{feynartspicture}
}
\hspace*{8em}
{\scriptsize \unitlength .6pt
\begin{picture}(210,125)(10,-15)
\put(-320,100){\small Full diagram:}
\put(-30,100){\small Subdiagram:}
\SetScale{.6}
\Line(20, 85)( 60, 85)
\Line(20,  5)( 60,  5)
\Line(140, 85)(180, 85)
\Line(180,  5)(140,  5)
\SetColor{PineGreen}
\Line(60, 85)(140, 85)
\Line(60,  5)(140,  5)
\Line(60, 85)( 60,  5)
\Line(140,  5)(140, 85)
\SetColor{Black}
\Vertex( 60, 85){3}
\Vertex( 60,  5){3}
\Vertex(140,  5){3}
\Vertex(140, 85){3}
\put(64,40){$0$}
\put(92,11){$\MZ$}
\put(95,67){$\Mu$}
\put(126,40){$0$}
\put(-5,80){$\Mu^2$}
\put( 6,0){$s$}
\put(185,80){$s_{\mu\bar\nu}$}
\put(185,0){$0$}
\CArc(100,50)(50,60,120)
\put(90,110){$s_{\mu\bar\nu\Pu}$}
\CArc(110,45)(50,-30,30)
\put(165,40){$s_{\mu\bar\nu\Pd}$}
\end{picture} }
\\[-1em]
\includegraphics[scale=0.7]{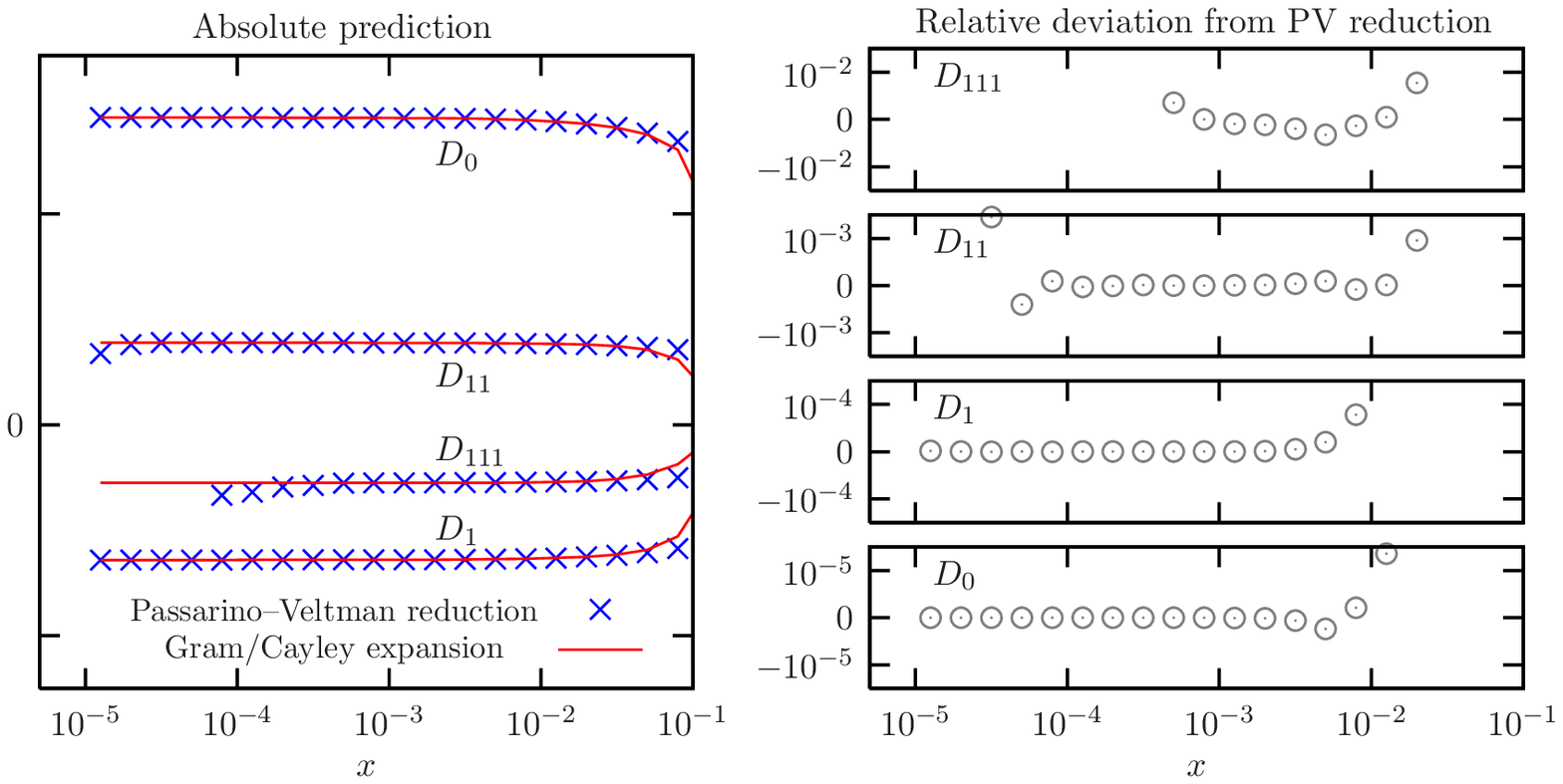}
\raisebox{3em}{\unitlength 1cm
\begin{picture}(3.7,4.4)
\put(0,4.2){\small Kinematics:}
\put(0.1,3.6){\scriptsize \parbox{.5cm}{$s$}$=4{\times}10^4\GeV^2$}
\put(0.1,3.2){\scriptsize \parbox{.5cm}{$s_{\mu\bar\nu}$}$=64{\times}10^2\GeV^2$}
\put(0.1,2.4){$|Z|,|X|\to0 \;\Leftrightarrow$}
\put(0.1,2.0){$s_{\mu\bar\nu\Pd}\to s\;$ and $\;s_{\mu\bar\nu\Pu}\to s_{\mu\bar\nu}$}
\put(0.1,1.2){\small Considered limit:}
\put(0.1,0.7){{$x\equiv {s_{\mu\bar\nu\Pd}}/{s}-1$}}
\put(0.1,0.2){{$\phantom{x} \equiv {s_{\mu\bar\nu\Pu}}/{s_{\mu\bar\nu}}-1\to0$}}
\end{picture} }
\vspace{-2.5em}
\caption{An example for 4-point integrals with both 
$|Z|$ and $|X|$ small ($x\to0$). Details as in Fig. \ref{fig:DsmallGram}.}
\label{fig:DsmallGramCayley}
\end{figure*}

Figure~\ref{fig:DsmallGram} illustrates the reliability of the
seminumerical method
for an example where the Gram determinant $|Z|$ becomes small
($x\to0$)
and therefore the   PV algorithm breaks down.
Away from the tricky region, i.e.\ for increasing $x$ (and thus $|Z|$),
there is a transition region in which PV reduction and the
seminumerical method agree within good accuracy until the latter
becomes less precise than the PV method.  This is due to the error of
the numerically evaluated coefficient $D_{0\dots0}$ which enters with
a prefactor $|Z|$, and is thus suppressed for small $|Z|$.

The seminumerical method is limited to cases in which $|X|$ is
not too small. If necessary, the case of both $|Z|$
and $|X|$ small, where PV reduction does not work either, can be cured 
with an expansion described below.
For 3-point integrals, the soft or collinear singular cases have the
property that $|X|=0$ exactly, so that the seminumerical method is not
applicable. However, these special cases are simple enough to be dealt
with analytically as described in App.~B of Ref.~\cite{Denner:2005nn}.

\subsection{\boldmath{Expansion for small $|Z|$}}

The PV relations can be rewritten as
\beqar
\lefteqn{
\Ymodadj_{0j} D_{i_1\ldots i_P} =
2\sum_{r=1}^P \Zadj_{ji_r} D_{00i_1\ldots \hat i_r\ldots i_P}}
\nn\\ && {}
+\Gramdet  D_{ji_1\ldots i_P}
+ \mbox{\small 3-point integrals},
\nn\\
\lefteqn{
\Zadj_{kl} D_{00i_1\ldots i_P}  \propto
\sum_{n,m=1}^{N-1} \Zadjadj_{(kn)(lm)}\Bigl[
f_nf_m D_{i_1\ldots i_P}}
\nn\\ && {}
+ 2 \sum_{r=1}^P (f_n\de_{mi_r}+f_m\de_{ni_r}) {D_{00i_1\ldots \hat i_r\ldots i_P}}
\nn\\ && {}
+ 4 \sum_{r,s=1\atop r\ne s}^P \de_{ni_r}\de_{mi_s} {D_{0000i_1\ldots
  \hat i_r\ldots \hat i_s\ldots i_P}}\Bigr]
\nn\\ && {}
+2m_0^2 \Zadj_{kl} D_{i_1\ldots i_P}
-\Gramdet D_{kli_1\ldots i_P}
\nn\\&&{}
+ \mbox{\small 3-point integrals},
\eeqar
which express the coefficients $D_{i_1\ldots i_P}$ and
$D_{00i_1\ldots i_P}$ in terms of lower-rank 4- and 3-point coefficients
and higher-rank 4-point coefficients dressed with a prefactor $|Z|$.
For small $|Z|$, these relations can, thus, be used for an iterative
determination of all 4-point coefficients from 3-point coefficients
up to terms that are suppressed by some powers of $|Z|$.
For a 
specific coefficient the iteration can be made arbitrarily precise upon
including higher and higher ranks in the iteration.

Figure~\ref{fig:DsmallGram} demonstrates the reliability of the method
for the same configuration as 
for the seminumerical method of the
previous section. For increasing $|Z|$ (and $x$) the expansion becomes
less and less precise, because the missing terms that are suppressed by
some power of $|Z|$ grow.
Again there is a region where all methods yield decent results.

The discussed expansion is limited to the case where
$\Ymodadj_{0j}$ and $\Zadj_{kl}$ are not too small for at
least one set of indices $j$, $k$, $l$. If all $\Ymodadj_{0j}$ are small,
then $|X|$ is small, too. This case is considered in the next subsection.
The case in which all $\Zadj_{kl}$ are small is further elaborated in
Ref.~\cite{Denner:2005nn}.

\subsection{\boldmath{Expansion for small $|Z|$ and $|X|$}}

The PV relations can again be rewritten as
\beqar
\lefteqn{
\sum_{r=1}^P \Zadj_{k i_r} 
{D_{00i_1\ldots \hat i_r\ldots i_P}}
\propto
{\Ymodadj_{k0} D_{i_1\ldots i_P}}
-{\Gramdet  D_{ki_1\ldots i_P}}}
\nn\\ && {}
+ \mbox{\small 3-point integrals},
\nn\\
\lefteqn{
 \Ymodadj_{ij} D_{i_1\ldots i_P}
= \mbox{\small const.}\times\Zadj_{ij} D_{00i_1\ldots i_P}
}
\nn\\ && {}
-2  \sum_{r=1}^P\sum_{n=1}^{N-1}\Zadjadj_{(in)(ji_r)}f_n
D_{00i_1\ldots \hat i_r\ldots i_P}
\nn\\
&& {}
+ \Ymodadj_{0j} D_{i i_1\ldots i_P}
+ \mbox{\small 3-point integrals},
\eeqar
which expresses the coefficients $D_{i_1\ldots i_P}$ and
$D_{00i_1\ldots i_P}$ in terms of lower-rank 4- and 3-point coefficients
and higher-rank 4-point coefficients dressed with a prefactor 
$|Z|$ or $\Ymodadj_{0j}$.
If $|Z|$ and all $\Ymodadj_{0j}$ are small, these relations can be
used to iteratively determine all 4-point from 3-point coefficients
up to suppressed terms. Compared to the case of the previous subsection,
where only $|Z|$ was considered small, the tensor rank grows faster
with each iteration.

Figure~\ref{fig:DsmallGramCayley} shows the reliability of the method
for a specific example, revealing again the breakdown of PV reduction
for sufficiently small $x$ (and thus $|Z|$ and $|X|$), 
while the expansion becomes less precise for increasing $x$.

The expansion method fails if either all $\Zadj_{k i_r}$ or all 
$\Ymodadj_{ij}$ are small. Possible treatments of these exceptional cases
are also described in Ref.~\cite{Denner:2005nn}.

\end{document}